\documentclass[twocolumn,10pt,fleqn]{revtex4}

\usepackage{txfonts}
\usepackage{mathrsfs}
\usepackage{paralist}
\usepackage{hyperref}
\usepackage{graphicx}
\usepackage{natbib}  

\begin{document}

\title{Testing DM halos using rotation curves and lensing: \\ 
A warning on the determination of the halo mass.}
\author{Dar\'io N\'u\~nez$^{1}$, Alma X. Gonz\'alez-Morales$^{1}$, Jorge L. Cervantes-Cota$^{2,3}$,
Tonatiuh Matos$^{4}$}
\affiliation{$1$Instituto de Ciencias Nucleares, Universidad Nacional
Aut\'onoma de M\'exico, A.P. 70-543,  04510 M\'exico D.F.,
M\'exico,}
\affiliation{$^2$ Depto. de F\'{\i}sica, Instituto Nacional de
Investigaciones Nucleares, M\'{e}xico,}
\affiliation{$^3$ Berkeley Center for Cosmological Physics, University of California, Berkeley, CA,
USA}
\affiliation{$^4$  Departamento de F\'isica, Centro de Investigaci\'on y de Estudios Avanzados del
IPN, A.P. 14-740, 
07000 M\'exico D.F., M\'exico,}

\collaboration{Instituto Avanzado de Cosmolog\'{\i}a, IAC} \noaffiliation

\email{nunez@nucleares.unam.mx}
\email{alma.gonzalez@nucleares.unam.mx}
\email{jlcervantes@lbl.gov}
\email{tmatos@fis.cinvestav.mx}

\begin{abstract}
There are two observations of galaxies that can offer some insight into the nature of the dark matter (DM), namely the rotation curves and the gravitational lensing. While the first one can be studied using the Newtonian limit, the second one is completely relativistic. Each one separately can not determine the nature of DM, but both together give us key information about this open problem.   In this work we use a static and spherically symmetric metric to model the DM halo in a galaxy or in a galaxy cluster. The  
metric contains two free functions, one  associated with the distribution of 
mass and the other one with the gravitational  potential.   
We use galactic, typical rotation curves to univocally determine the kinematics of 
the halos.  We compute separately the mass functions for a perfect fluid and a scalar field,
and demonstrate that both models can be fitted to the observations, though with different
masses. We then employ lensing  to discriminate between these  models. This
procedure represents a test of models using two measurements: rotation curves and lensing.   
\end{abstract}

\date{\today}

\maketitle

There are at least four main groups of cosmological observations which are consistently described by
the $\Lambda$CDM model, namely: the acoustic peaks in the cosmic microwave background radiation,
Supernovae type Ia data, rotation curves of spirals and dynamics of galactic clusters, and their
gravitational lensing \citep{Spergel:2006hy,Tegmark:2006az,Riess:2006fw}.  The last two of these
observations are related most significantly with
the presence of dark matter (DM), where a consistent model considers a DM halo surrounding the
galaxies and galactic clusters.

Using Newtonian gravity we can infer the mass of the DM halos from observations of velocity profiles
in galaxies and galaxies clusters. Alternatively, one can use lensing images or distortions of background
galaxies due to the space-time curvature \citep{Dye:2007nv,2003AIPC..666..113N}.  This  is 
notorious, particularly in the case of gravitational lensing, which is a general relativistic 
effect {\it per excellence}, as it strictly can not be dealt with using Newtonian physics.

There has been many attempts to model DM halos in the general relativistic framework, some
of them has shown that general relativistic effects can be important
\citep{1997apj..482..963N,1998PhRvD..58h3506N,Matos:2000jr,
Guzman:2000zba,CervantesCota:2009my,Nandi:2009hw,Bharadwaj:2003iw,Faber:2005xc}, but in general
they make use of approximations or limits that in some cases represents assumptions about DM  properties.
The scalar field DM model is an alternative proposed in the past \citep{Matos:1999et} to fit the observed amount of 
substructure \citep{Matos:2000ss}, the critical mass of galaxies \citep{Alcubierre:2001ea}, the rotation curves of galaxies 
\citep{Boehmer:2007um}, the central density profile of  LSB galaxies \citep{Matos:2003pe}, the evolution of the cosmological 
densities \citep{Matos:2008ag}, among other topics.

In the present work we focus on two models for the DM halo (a perfect fluid and a scalar field) and
explicitly show what exactly is determined by the observations and what comes as extra assumptions.  This is particularly important
as the unknown nature of the DM is one of the most relevant questions that one would like to solve.
It is clear that to make assumptions specifically on the nature of DM, in order to obtain information on the 
nature of the DM, is skewing the problem.

In our study we consider a static and spherically symmetric space-time in General Relativity, 
described by the line element:
\begin{equation}
ds^2=-e^{2\Phi/c^2} \,c^2\,dt^2+ \frac{dr^2}{1-\frac{2\,G\,m}{c^2\,r}}+r^2 \, d\Omega^2,
\label{eq:lel}
\end{equation}
where $d\Omega=d\theta^2+\sin^2\theta d\varphi^2$. The gravitational potential $\Phi(r)$ and the
mass function $m(r)$ are functions of the
radial coordinate only. In fact, due to the symmetries of this space-time, all physical quantities depend
only on $r$. The Einstein's equations are the known set \citep{1985fcgr.book.....S}:
\begin{eqnarray}
m'&=&- \frac{4 \pi r^{2}}{c^{2}} {T^t}_t, \label{eq:einstein00} \\
\left(1-2\,\frac{m\,G}{c^{2} r}\right)\,\frac{\Phi'}{c^{2}} -
\frac{m\,G}{c^{2} r^2} &=&\frac{4\pi\,G\,r}{c^{4} }  {T^r}_r ,
\label{eq:einstein0}
\end{eqnarray}  
where prime $' \equiv \partial/{\partial r}$. The potentials of such space-time are determined by 
the DM halo. In order to see the difference between two hypothesis on the nature of DM, we will consider 
two types of composition for halos:  a perfect fluid and a scalar field.

The  above equations are complemented by the conservation equation of the matter-energy generating 
the curvature of the space-time.  But given the different nature of the fluids considered here, this
equation is treated separately for each fluid. 

In the case of the perfect fluid, the stress-energy tensor is given by 
$ T_{\mu\,\nu}=(\rho +p/c^{2} ) u_{\mu} u_{\nu} + p g_{\mu\,\nu} $, 
where the density is $\rho=(1+\epsilon)\rho_{0}$, where $\rho_{0}$ is the rest mass energy
density and $\epsilon$ the internal energy per unit mass, $u^{\mu}$ is the co-movil 
four velocity, normalized as $u_{\mu}u^{\mu}=-c^{2}$, and $p$ is the pressure.   The conservation 
equation, ${T^\mu}_{\nu ; \mu}=0$, implies the field equation:
\begin{equation} \label{contpf}
\left( \rho c^{2} + p \right) \frac{\Phi'}{c^{2}} + p'=0,
\end{equation}
which can be rewritten as
\begin{equation}
{{T^{r}}_{r}}'+\left(T^{r}_{r}- T^{t}_{t} \right) \frac{\Phi'}{c^{2}} = 0 . \label{eq:pf0}
\end{equation}

Now, for the scalar field the stress energy tensor is given by 
\begin{equation}
{T_{\mu\,\nu}} = \phi_{,\mu}\,{\phi_{,\nu}} - 
\frac{1}{2}\, g_{\mu\,\nu}\left( g^{\alpha \beta}\phi_{,\alpha}\,\phi_{,\beta}+ 2\,V\left(\phi\right)
\right) ,
\end{equation}
where $\phi_{,\alpha}=\partial \phi/ \partial x^{\alpha}$; and $V\left(\phi \right)$ is the scalar
potential. The  components of the stress-energy tensor are
\begin{eqnarray}
 {T^{t}}_{t}&=& - \frac{1}{2}\left(1- \frac{2\,
G\,m}{c^2\,r}\right){\phi'}^2 - V\left(\phi\right) , \nonumber \\
 {T^{r}}_{r}&=&\frac{1}{2}\left(1- \frac{2\,
G\,m}{c^2\,r}\right){\phi'}^2- V\left(\phi\right)  \label{eq:tmn_sf} ,\\
{T^{\theta}}_{\theta}&=&{T^{\varphi}}_{\varphi}={T^{t}}_{t} . \nonumber
\end{eqnarray}

From the conservation equation for the scalar field, ${T^\mu}_{\nu ;\mu} = 0$, one 
obtains a field equation, the Klein-Gordon equation,
\begin{equation}
\phi''+
\left(  \frac{\Phi'}{c^{2}} -  \frac{\frac{m'\,G}{c^2\,r} +
\frac{\frac{3\,m\,G}{c^2}-2\,r}{r^2}}{1-\frac{2\,m\,G}{c^2\,r}}
\right)  \phi' - \frac{\frac{\partial V}{\partial \phi}}{1-\frac{2\,m\,G
}{c^2\,r}}=0, \label{eq:ecsf}
\end{equation}
that can be written as
\begin{equation}
 {{T^{r}}_{r}}'+\left(T^{r}_{r}- T^{t}_{t} \right)\left(\frac{\Phi'}{c^{2}} + \frac{2}{r}\right)=0,
\label{eq:sf0}
\end{equation}
which is remarkable similar to the field equation for the perfect fluid, Eq.(\ref{eq:pf0}). Given 
this similarity, it is convenient for our mathematical description to consider the single
field equation for both types of matter
\begin{equation}
 {{T^{r}}_{r}}'+\left(T^{r}_{r}- T^{t}_{t} \right) \left(\frac{\Phi'}{c^2}+\frac{2 a}{r}\right)=0.
\label{eq:pfandsf}
\end{equation}
in which $a=0$ for the perfect fluid, and $a=1$ for the scalar field. Notice that if one considers
a sort of perfect fluid given by ${T^\mu}_\nu={\rm diag}(-\rho,p,p_i,p_i)$, $p_i$ (some times
called ``tangential'' pressure) denotes a
term representing the ignorance we have on the features of the fluid. This presure $p_i$ is related to the other fluid
variables as $p_i=\left(1-a\right)\,p-a\,\rho c^{2}$, (see 
Eq.~(\ref{eq:pfandsf})), where ``$a$'' takes, in principle, any value. There are works which have
discussed this field equation considering $a$ as a free parameter
\citep{Faber:2005xc,Nandi:2009hw,Bharadwaj:2003iw}.
For the purpose of the present work we will consider only the two extremal cases, $a=0$ and $a=1$,
but the discussion can be directly applied for these cases as well.

In this way, the system of equations which must be solved are, the Einstein's equations,
Eqs.~(\ref{eq:einstein00},\ref{eq:einstein0}), and the field equation, Eq.~(\ref{eq:pfandsf}). In
either case, there are four unknown functions, $m, \Phi, p$ and
$\rho$, for the case where the curvature of the space-time is due to the perfect fluid, or $m, \Phi,
\phi$ and $V(\phi)$ when the curvature is caused by the scalar field. Thus, we have three equations
for four unknown functions. In either case, we need only one extra data. It is important to underline this
fact. Once the extra data
is given, there is no more room left for any other assumption, the rest of the functions are
determined by the system of equations. If, for instance, we give an equation of state for the
perfect fluid, $p=p(\rho)$ or, in the case of the scalar field, an explicit form for the potential,
$V(\phi)$, there is no freedom left to choose the form of the rest of the functions, they will be
determined by the system of equations.

Following the line of work presented in \citep{Matos:2000jr}, we use  observational results to close
the system of equations.  In the case of galactic halos, two main observations can serve to obtain
the desired information: measurements of rotation curves in spirals and light deflection by lensing.
In this work we choose the former to complement the above field equations and use the latter to
discriminate between different halo type models.

{\it Rotation curves}.  The motion of test particles in such spacetime is determined by the
geodesic equations and, for test particles in circular motion, there is a relationship between the
gravitational potential, $\Phi$, and the tangential velocity of those particles, $v_c$:
\begin{equation}
\frac{\Phi'}{c^2}=\frac{\beta^2}{r} \label{eq:phi_v}  , 
\end{equation}
where we have defined $\beta^2=\frac{{v_c}^2}{c^2}$. This tangential velocity
is the one measured by  observations of rotation curves in galaxies. Thus, $v_c$ is an observable 
function, and by means of Eq.~(\ref{eq:phi_v}), the gravitational 
potential can be determined. Thus, given this observable, there is no room left for an equation of state for the
perfect fluid or for a given scalar field potential.

Moreover, as long as the magnitude of the observed velocities are small with respect to the speed
of light, this justifies the validity of one of the weak field approximations $\Phi/c^2 << 1$ that
one usually assumes by taking the weak field limit. 
Here we want to emphasize that the approximations $2\,G\, m\, /\,c^2\,
r\,<<1\,$ and especially  $p<<\rho$ are, in general, extra hypothesis which strongly depend upon the nature of the
DM type. It is clear that if all these conditions are satisfied, then the above system of
equations, Eqs. (\ref{eq:einstein00}, \ref{eq:einstein0}, \ref{contpf}), 
reduces to the hydrodynamic set of equations for the case of the perfect fluid model. But, for example in the case of the scalar
field, there is no Newtonian limit, and one has to be careful with these approximations.

After substituting Eq.~(\ref{eq:phi_v}) into the gravity equations, 
Eqs. (\ref{eq:einstein00}, \ref{eq:einstein0}), we obtain an equation (with no approximations) for 
the mass function as the only free function
\begin{equation}
m'+P(r)\,m = Q(r)
\end{equation}
with
\begin{eqnarray}
P(r)&=&\frac{2\,r\,{\beta^2}' -\left(1+2\,\beta^2\right)\,\left(3-2\,a-\beta^2\right)}{
\left(1-2a+\beta^2\right)\,r} ,\nonumber \\
Q(r)&=&\frac{c^2}{G}\frac{r\,{\beta^2}'- \beta^2\,\left(2-2\,a-\beta^2\right)}
{\left(1-2a+\beta^2\right)} .
\end{eqnarray}
The functions  $P(r)$ and $Q(r)$ depend on the type of fluid we are dealing with ($a$) and on the rotation
curves profile.

The mass function can be expressed in terms of the gravitational potential, $\Phi$, through the
integral 

\begin{equation}
m=\frac{\int{e^{\int^r{P(r')dr'}} Q(r) dr + C}}{e^{\int^r{P(r')dr'}}}.
\label{eq:mpfsf}
\end{equation}
where the value of the integration constant, $C$, is set by the appropriate boundary conditions.   

For the case of the perfect fluid, the density and pressure are directly computed from Eqs.
(\ref{eq:einstein00}) and  (\ref{eq:einstein0}), respectively.  

For the scalar field, using the expressions  Eqs.~(\ref{eq:tmn_sf}), we obtain that 
\begin{eqnarray}
{\phi'}^2&=& \frac{c^4}{4\,\pi\,G\,r^2}\left(\frac{\frac{G\,m'}{c^2}-\frac{G\, m}{c^2\,
r}}{1-\frac{2\, G\,m}{c^2\, r}}+\beta^2\right) , \label{eq:chi} \\
V(\phi(r))&=&\frac{c^2}{8\,\pi\,r^3}\left[m\,\left(1+2\, \beta^2\right)+m'\,r\right]\nonumber\\
&-&\frac{c^4}{8\,\pi\,G}\left(\frac{\beta^2}{r^2}\right).
\label{eq:Vchi}
\end{eqnarray}

Once the function $\beta(r)$ is given,  the scalar field and its potential are straightforwardly 
determined in terms of the radial coordinate. In order to obtain the form $V(\phi)$, one needs to
invert the solution for the scalar field ($r=r(\phi)$), and to substitute it into
Eq.~(\ref{eq:Vchi}). As shown below, this procedure works at least for simple $\beta$ functions.

In this way, we have shown that the mass function $m$, associated to a galactic halo by means
of the rotation velocity strongly depends on the DM model that is being considered. The
single observation of the rotation curve is not sufficient to determine the nature of DM
and hence the mass associated with the halo. Moreover, we have shown that the relationship between the
pressure and the density, or between the scalar field and the scalar potential is fixed, up to
integration constants, once the rotation velocity is employed.

{\it Lensing}.  The other observation concerns the gravitational lensing, that for the line element 
given by Eq.~(\ref{eq:lel}), the deflection of the light ray, $\Delta\,\varphi$, at the radius of
maximal approach, $r_m$, is given by  \citep{2002glml.book.....M},
\begin{equation}
\Delta\,\varphi = -\int_{\infty}^{r_m}
\frac{r_m\,dr}{r^2 \sqrt{\left(1-\frac{2\,G\,m}{r c^{2}}\right)
\left[e^{-2\frac{ \Phi}{c^2}}e^{2\frac{\Phi(r_m)}{c^{2}}}-\frac{r_m^2}{r^2}\right]}}.
\label{eq:deflangle}
\end{equation}

Since the gravitational potentials and the fluid variables are already determined by the rotation
curves of spirals, deflection angle measurements can serve to discriminate between models. Here 
we deal with two examples,  perfect fluid and scalar field DM models. Yet, observations of spirals that 
lenses light are not very common, however the first examples of them has  recently appeared
\citep{2010MNRAS.401.1540T}.  
We remark that from the expression of the deflection angle, it is a large step to infer the
mass of the DM halo based solely on the observation of the deflection angle. A supposition
has to be made on the relation between the gravitational potential, $\Phi$, and the mass function,
$m$ \citep{2002glml.book.....M}. Such supposition, as we have shown, not only strongly
depends on the type of matter, but also on the specific characteristics of the type of matter
considered. 

In the next section we present some examples of known rotation curve profiles to give a quantitative
description to these conclusions.

{\it Examples}.  The idea in this section is to stress the conclusions that we are presenting by
means of considering a typical observation of rotation velocities in spirals and to directly
determine the gravitational mass in each case,  when the DM is a perfect fluid (dust)  
and when it is a scalar field. 

In practice we can consider a velocity distribution, as a phenomenological model, 
for instance the velocity profile coming from N-body simulations given by 
NFW  \citep{Navarro:1995iw,Navarro:1996gj} or a Burkert profile \citep{Burkert:1995yz} 
given by the phenomenological of rotation curves \citep{Salucci:2007tm}, to determine the 
mass of  each type of fluid.   We will show  that the gravitational mass inferred by the same 
velocity profile is strikingly different for the perfect fluid and scalar field cases.

{\it Constant velocity profile.}
We will consider the simplest case of constant rotation curves as our first example. 
Although there are some examples of galaxies that present a constant 
velocity profile, for a few disk length scales  \citep{Sofue:2000jx}, this is not a typical behavior, being our 
own Galaxy a good counter example \citep{Gnedin:2010fv} and, in fact, there is an important 
rotation curve phenomenology described by the Universal Rotation Curve \citep{Salucci:2007tm,Salucci:2009yp}.
However, the constant velocity profile offers us the mathematical simplicity to obtain straightforward analytical results 
and to show the main point of our work. 

For the gravitational potential, from Eq.~(\ref{eq:phi_v}), when the velocity function is a
constant, $\beta_0$, we get
\begin{equation}
\Phi=c^2\,\ln \left(\frac{r}{r_0} \right)^{{\beta_0}^2}. \label{eq:phi_v_cte}
\end{equation}
The mass function can be analytically obtained for any value of the parameter $a$ as:
\begin{eqnarray}
 m_{\beta_0}=&&\frac{c^2}{G}\,\left(\frac{{\beta_0}^2\,\left(2\,\left(1-a\right) -
{\beta_0}^2\right)}{2\,\left(1+2\,\left(1-a\right){ \beta_0}^2 - {\beta_0}^4\right)}\,r + \right.
\nonumber \\ 
&& \left. C\,r^\frac{\left(1+2\,{\beta_0}^2\right)\,\left(3-2\,a-{\beta_0}^2\right)}{1-2\,a +
{\beta_0}^2}\right), \label {eq:m0_a}
\end{eqnarray}
where $C$ is the integration constant of Eq.~(\ref{eq:mpfsf}). For the case of DM
described by a perfect fluid, $a=0$, we fix this constant to zero in order to avoid changes in the
signature of the line element, Eq.~(\ref{eq:lel}). Thus, the mass function, and the corresponding
pressure and density in the case of the perfect fluid are given by
\begin{eqnarray}
m_{pf}&=&\frac{c^2}{2\,G}\,\frac{{\beta_0}^2\left(2-{\beta_0}^2\right)}{1+2\,{\beta_0}^2 -
{\beta_0}^4}\,r, \label{eq:m_pf_cte} \\
\rho&=&\frac{c^2}{4\,\pi\,G}\,\frac{{\beta_0}^2\left(2-{\beta_0}^2\right)}{r^2\left(1+2\,{\beta_0}^2
-{\beta_0}^4\right)},  \\
p&=&\frac{c^4}{8\,\pi\,G}\,\frac{{\beta_0}^4}{r^2\left(1+2\,{\beta_0}^2 -
{\beta_0}^4\right)}, \\
\frac{p}{\rho} &=&  \frac{\beta_0^2 \, c^{2}}{2(2-\beta_0^2)} = {\rm const.}
\end{eqnarray}
We can see in the limit of very small velocities, $\beta_0<<1$, we recover the Newtonian limit, and
pressure is negligible with respect to the density as we mentioned above, however, it is not zero
and, actually, we obtain a barotropic equation of state, $p = w_0\,\rho$.

On the other hand, considering the DM halo due to a scalar field, the mass function is
obtained from Eq.~(\ref{eq:m0_a}), with $a=1$. In this case, the mass function has a very
peculiar behavior. The first term is small, proportional to ${\beta_0}^4$, but
negative. The second term, proportional to the constant $C$, goes as
$r^{-\left(1+2\,{\beta_0}^2\right)}$, thus, by choosing a positive value for the constant $C$, 
one can have a positive mass function for a large region, but this function will present a
divergence at the origin. Of course this result was expected, as the space-time metric is 
static. In order to avoid this problem, we had to take non-static space-times, like the oscillatons
\citep{Alcubierre:2001ea}, but 
this is beyond the scope of this work.  It can be shown, however, that this divergence is covered by an apparent
horizon. Some features of this case of scalar field with a non zero constant $C$ in the mass
function, have been discussed in \citep{Nandi:2009hw}. For the purpose of this work, we only notice
that the geometric functions and those of the scalar field, have a non intuitive behavior, but are
consistent with the rotational curve. Explicitly, for the case of $C=0$, the mass function is
\begin{equation}
m_{sf}=-\frac{c^2}{2\,G}\frac{{\beta_0}^4}{1-{\beta_0}^4}\,r, \label{eq:m_sf_vcte} 
\end{equation}
and, with the geometric functions determined, the scalar field and scalar potential are
completely fixed, given by:
\begin{eqnarray}
\phi&=&\pm\sqrt{\frac{c^4}{4\,\pi\,G}}\,{\beta_0}\,\ln\left(\frac{r}{r_0}\right) \\
V(r)&=&-\frac{c^4}{8\,\pi\,G}\,\frac{{\beta_0}^2}{r^2\,\left(1-{\beta_0}^2\right)} \\
V(\phi)&=&
-\frac{c^4}{8\,\pi\,G}\,\frac{2\,{\beta_0}^2}{\left(1-{\beta_0}^2\right)\,{r_0}^2}\,e^{\mp\,2\,
\sqrt {\frac{4\,\pi\,G}{c^4}}\,\frac{\phi}{\beta_0}}.
\end{eqnarray}
where the expression for the scalar field, $\phi(r)$, was inverted to obtain $r(\phi)$, and then
express the scalar potential in terms of $\phi$, as explained previously. The  ``effective mass'' of the 
scalar field,   
$m_{\rm eff}\sim \frac{2}{\sqrt{1-{\beta_0}^2}\,r_0}$, depends inversely on
the characteristic distance of the halo. This distance is of the order of kilo-parsecs, and 
$\beta_0\sim 10^{-3}$, thus it will turn into a typical mass for scalar field in a galaxy, which corresponds 
a very light boson mass $\sim 10^{-23}$eV$/c^2$. This result is in agreement with the one
obtained in previous works, see  for example \citep{Matos:2000ss}.

Going back to our previous discussion, notice how remarkably different are the mass expressions
derived from each type of fluid, being both
consistent with the observed rotation velocities. This is the simplest case in which we can show how
the single observation of the rotation velocities in halos determines the features of the
perfect fluid model or the scalar field.

Although the mass associated to the scalar field results 
negative and this can be taken as a no-go result for static scalar field halos 
\citep{DiezTejedor:2006qh}, rotation curves of spirals are not exactly flat (see discussion in 
 \citep{Salucci:2007tm,Salucci:2009yp}) and, in addition, we have to be cautious with the supposition of a static  
metric which is very restrictive for the scalar field . Thus, the above-result  
should not be taken as definitive, at most, it should be taken as a remark that a static DM halo is not   
well described by a static scalar field.   A negative mass, or positive gravitational potential, is known since long
time ago \citep{1984A&A...136L..21S} from the fits to rotation curves using modifications of
newtonian gravity in which a scalar field induces a  Yukawa--type force.  At the end, demanding a
constant velocity profile all the way in the radial direction  implies an effective repulsive force
to be acting on test particles in the galaxy.

In any case, it emphasizes our point in showing how strongly depends the determination of the mass 
of the DM halo on the type of matter considered to describe it.

The deflection angle for the case of constant rotation velocity, considering the perfect
fluid and the scalar field with the constant $C=0$, implies the following
expressions:
\begin{eqnarray}
\Delta\,\varphi &=& \int_{0}^{1}{
\frac{dx}{\sqrt{A_{type} \,\left(x^{2{\beta_{0}^2}}-x^2\right)}}}, \\
A_{pf}&=& \frac{1}{1-{\beta_0}^2 \left({\beta_0}^2-2\right)}
\label{eq:deflangle_pf} \\
A_{sf}&=&\frac{1}{1 - {\beta_0}^4}, 
\label{eq:deflangle_sf} 
\end{eqnarray}
where we have defined $x=\frac{r_m}{r}$. Since in any case the deflection angle is a constant,
i.e.
it does not depend of the maximal approach radio $r_m$, it can be evaluated for a typical value of
the velocity. For comparison we take the value $\beta_0=1/1200$, that corresponds to a velocity of
$v_c=250\,$ km/s. Evaluating the deflection angle, we get
\begin{eqnarray}
\alpha_{pf}= 0.899547, \\
\alpha_{sf}= 0.449546,
\end{eqnarray}
for both cases the deflection angle is given in arc second units. We can see that there is a
difference of almost half arc second between them, and the simultaneous observation of the rotation
velocity and the deflection of light produced by the galactic halo, can teach us about the true
nature of the DM. 

We notice that the deflection angle for the scalar field with a non-zero value of the constant $C$
in the mass function takes very large values, a fact which certainly allows us to discard this
option as a model for the DM halo, independently of any interpretations of the mass
function. 

 Now we study an example that is less striking though.

{\it Navarro-Frenk-White (NFW) velocity profile.}  
Independently from its origin, the NFW profile \citep{Navarro:1995iw,Navarro:1996gj} 
 is considered {\it per se} as a viable fitting model to describe galactic kinematics. This 
profile has been subject to geometrical studies elsewhere
\citep{Matos:2004ev,Matos:2004je,Matos:2004ev}. 
In this example, we assume this profile as a valid phenomenological galactic profile for the
galactic data.  
We  obtain the usual expression for the mass derived within this description,
and compare it with the same form of the rotation velocity, but considering that it is due to a
DM halo composed of a scalar field. 

\begin{figure}[h]
\centering
\includegraphics[width=6cm]{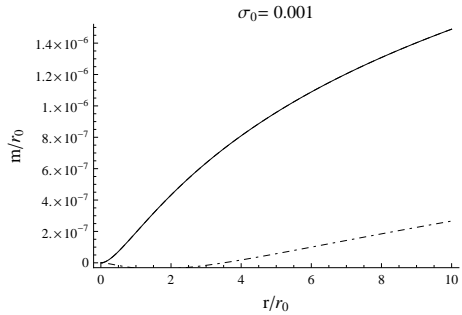}
\caption{Comparison of the masses using the rotation velocity profile from the NFW model with a
perfect fluid (upper curve) 
and scalar field (lower curve).}
\label{fig:m_nfwpfsf}
\end{figure}

\begin{figure}[h]
\centering
\includegraphics[width=6cm]{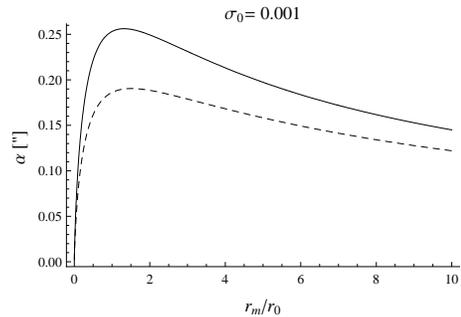}
\caption{Deflection angle generated by  the gravitational lensing of a NFW rotation profile with a
perfect fluid (upper curve) 
and scalar field (lower curve). }
\label{fig:def_nfw}
\end{figure}

The velocity profile in the NFW model \citep{Navarro:1995iw,Navarro:1996gj} is given by
\begin{equation}
v_T^{2}=\frac{{\sigma_{0}}^2 r_{0}}{r} \left(-\frac{r/r_{0}}{1 + r/r_{0}} + \ln\left[1 +
r/r_{0}\right]\right).
\label{eq:v_nfw}
\end{equation}
where $\sigma_{0}= 4\pi\,G\,\rho_0\,{r_{0}}^2$ is a characteristic velocity of stars in the halo, 
given in terms of a characteristic density, and $r_0$ is a scale radius. Given this velocity
profile, we have to solve  Eq. (\ref{eq:mpfsf}) with $a=0$ for the perfect fluid and  with $a=1$ for the 
scalar field.  In neither case there is an analytical solution, thus we have integrated the equations
numerically.  We do not want to treat here specific galaxies but to emphasize the differences
between the galaxy models. Therefore, we set $\sigma_{0} $ and $r_0$ to some typical values. 
In our plots we assume  geometric units ($G = c = 1$), and therefore the characteristic velocity 
takes values, $0<\sigma_{0}<1$, and the mass is less than the unity. For definiteness,
we assume $\sigma_{0} = 0.001$ and $r_0=1$. In figure (\ref{fig:m_nfwpfsf}) we plot both halo masses
(perfect fluid and scalar field). Disregarding the behavior near the origin, as long as we are
considering the outside region, as mentioned above, we see that the mass associated to the halo in
each case are different. 
We now consider lensing.  By integrating  Eq. (\ref{eq:phi_v}) for the given rotation velocity, Eq.
(\ref{eq:v_nfw}), we obtain the following expression for the gravitational potential:
\begin{equation}
\Phi=- \sigma_{0}\,r_0\frac{\ln\left(1+\frac{r}{r_0}\right)}{r}.
\end{equation}
We substitute this expression, together with the corresponding numerical solution one for the mass
in each case, in the
equation for the deflection angle, Eq.~(\ref{eq:deflangle}), and perform the integration varying
the value of the radius of maximal approach, $r_m$.  The results are plotted in figure
\ref{fig:def_nfw}. As we see, the observation of the deflection angle can determine which type of matter  
is actually composing the DM halo.

In this example, the mass associated to the scalar field model is essentially positive and a
well--behaved function, that is, it does not follow the no-go result mentioned in the previous
example. Though there is a small region where the mass becomes negative. This is due to the fact
that we are demanding the velocity profile to grow in that region. In a real setting however the
stellar disc  adds to the velocity profile, thus we expect that  its contribution avoids negative
mass regions for the scalar field.

With these examples it is clear how two different types of matter (perfect fluid and scalar field) can
be consistent with the observation of rotation curves of DM halos, though they lead to
different conclusions to the mass function inferred. The deflection of light can then be used to
discriminate between the two models. Even though the mass function for some model has not an
intuitively expected behavior, it is necessary to use the observation in order to discard the
model, being aware of the assumptions made during the derivation of such conclusions.

\vspace{-1.0cm}

\begin{acknowledgments}
This work was supported by CONACYT Grant No. 84133-F and  UC MEXUS-CONACYT Visiting Scholar 
Fellowship Program grant for JLCC.
\end{acknowledgments}

\bibliography{nunez_etal_corrected2} 

\end{document}